\documentclass{aa}
\usepackage{times,graphicx}
\usepackage{supertabular}
\usepackage{curves,epic,eepic}
\usepackage{float}

\begin{document}

\thesaurus{03.13.5; % A&A Section 6: Form. struct. and evolut. of stars
           13.07.1;} %Gamma rays: bursts,

            \title{Optical observations of GRB 
                afterglows: GRB 970508 and GRB 980326 revisited
                ~\thanks{Based on observations
                collected at the German-Spanish Astronomical Centre, Calar
                Alto, operated by the Max-Planck-Institut f\"ur Astronomie,
                Heidelberg, jointly with the Spanish National Commission
                for Astronomy.}$^{,}$\thanks{Based on observations carried
                out at the  4.2-m William Herschel Telescope operated
                by the Royal Greenwich Observatory in the Spanish 
                Observatorio del Roque de los Muchachos of the Instituto de 
                Astrof\'{\i}sica de Canarias.}}

   \author{A. J. Castro-Tirado
          \inst{1,2}
   \and J. Gorosabel
          \inst{1}}

   \offprints{J. Gorosabel (jgu@laeff.esa.es)}

\institute{Laboratorio de Astrof\'{\i}sica Espacial y F\'{\i}sica
     Fundamental (LAEFF-INTA), P.O. Box 50727, E-28080 Madrid, Spain.
\and Instituto de Astrof\'{\i}sica de Andaluc\'{\i}a (IAA-CSIC), P.O. Box
     03004, E-18080 Granada, Spain.}
         
   \date{Received date; accepted date}
   
   \titlerunning{GRB 970508 and GRB 980326 revisited}

   \authorrunning{Castro-Tirado and Gorosabel}

   \maketitle

   \begin{abstract} 
     Since January 1997, we have monitored 15 GRB fields, detecting 6 
     optical/IR afterglows.
     We have revisited GRB 970508 and GRB 980326. For GRB 970508, 
     we derive a power-law decay exponent $\alpha$ = -1.19 (R-band). 
     The luminosity of the host galaxy $L$ relative 
     to the characteristic luminosity $L^{\star}$ is in the range 0.06--0.15,
     i.e. a dwarf galaxy.
     For GRB 980326, we derive a power-law decay exponent $\alpha$ = -1.7, 
     taking into account the new upper limit for the host as R $\geq$ 27.3
     provided by Bloom and Kulkarni (1998). This implies one of the fastest 
     GRB optical decays ever measured.
     The fact that only about 50 \% of optical transients have been 
     found within the $\gamma$/X-ray error boxes, suggest that
     either considerable intrinsic absorption is present  
     or that some optical transients display a very fast decline.
     We also propose that the $^{\prime\prime}$secondary 
     maximum$^{\prime\prime}$ detected on 17 Apr 1998 could be explained in 
     the context of the $^{\prime\prime}$SN-like$^{\prime\prime}$ light
     curves 2-3 weeks after the GRB, as recently suggested by Woosley (1999).
   \end{abstract}

%
%  14.Sep.'90: Demo-Vs.
%________________________________________________________________

\section{Introduction}

Since 1997, we have been monitoring the error boxes of gamma-ray bursts
(hereafter GRBs) detected by the BSAX or RXTE satellites. 
We have monitored 15 GRB fields on the aggregate, detecting 6 
optical/IR afterglows. 
See Castro-Tirado (1999) and Gorosabel (1999) for a review of results. 
Here we focus on the GRB 970508 and GRB 980326 afterglows. 

%The telescopes at Calar Alto are well suited
%for this task, because of the high number of clear nights with good
%observing conditions and flexibility. 
%Optical imaging (and eventually spectroscopy) can be performed with a 
%variety of CCDs at the Calar Alto (1.23-m, 2.2-m and 3.5-m). Infrared 
%imaging is possible with the MAGIC and OMEGA detectors at Calar Alto.

%See Table 1 for a summary.

%\begin{table}
%\begin{center}
%\caption{Log of optical/IR follow-up observations obtained at CAHA.}
%\begin{tabular}{|c|c|c|c|c|} \hline
%   GRB         &  Telescope    & Reference \\
%\hline
%GRB 970111     &   2.2m        &   1,2     \\
%GRB 970228     &   2.2m        &           \\
%GRB 970508     &   2.2m        &   3-5     \\
%GRB 970616     &   3.5m        &   6,7     \\
%GRB 970815     &   2.2m,3.5m   &    8      \\
%GRB 970828     &               &           \\
%GRB 971214     &   3.5m        &    9      \\
%GRB 971227     &   2.2m        &  10,11    \\
%GRB 980326     &   2.2m        &   12      \\
%GRB 980613     &   1.2m,3.5m   &   13      \\
%GRB 980703     &   1.2m,3.5m   &   14      \\
%GRB 981220     &   1.2m,3.5m   &   15      \\
%GRB 981226     &   1.2m,3.5m   &   16      \\
%               &               &           \\
%\hline
%\end{tabular}
%\end{center}
%\end{table}

\section{Observations: some selected results}

\subsection{GRB 970508 and its host galaxy}

 The optical counterpart was discovered by Bond (1997).
 After a initial phase of constant brightness lasting one day (as
 inferred from the light curves shown in Fig. 1), a strong flare took place. 
 The peak brightness of the optical light occurred two days after the GRB,
 and a maximum was reached in the R and U bands at the same time. The 
 increase in brightness was similar to the one inferred for GRB
 970228 (Guarnieri et al. 1997).  

 New U and R-band images taken on 11 May showed that the source was 
 declining in brightness (Castro-Tirado et al. 1997) and by ten months
 later, it almost reached the quiescent level, revealing a 25th mag
 galaxy in the R-band (Bloom et al. 1998a, Castro-Tirado et al. 1998a). 
 The decay of the optical flux could be modelled in terms of a power law  
 decay, with $F_R \propto t^{-1.19 \pm 0.01}$
 ($\chi^2/dof=2.56$), assuming a host galaxy with R = 25.72 $\pm$ 0.20
 according to Bloom et al. (1998). On the basis of an extrapolation of 
 the V- and B-band measurements into the U-band, assuming a given spectral
 shape (Bloom et al. 1998b, Zharikov et al. 1998), we obtain U = 26.56 $\pm$
 0.25 for the host. If the U-band flux also followed a power-law decay, 
 then $F_U \propto t^{-1.69 \pm 0.34}$ ($\chi^2/dof=2.37$), implying a
 faster decay than that observed at longer wavelength. This 
 difference in decay slope between $F_R$ and $F_U$ is
 marginal given the large error on the U-band decay slope.

 With respect to the host galaxy, and depending on the value of the 
 $K$-correction and spectral index $\beta$ ($F_\nu \propto \nu^{-\beta}$) 
 considered, we obtain $M_{B}=-17.53
 \pm 0.5$ ($\beta=-3$) or $M_{B}=-16.6 \pm 0.6$ ($\beta=-1.56$). The first
 value is consistent with that reported by Zharikov et al. (1998). 
 In both cases the object is well below the knee of the galaxy luminosity
 function, $M^{\star}_{B} \sim -20.6$ (Schechter 1976) and the luminosity of 
 the host galaxy $L$ relative to the characteristic luminosity $L^{\star}$ is
 0.06 or 0.15 depending of the value of $M_{B}$ considered. 
 The host galaxy corresponds to any of the following categories: a
 starburst, a red dwarf starburst, a irregular dwarf, a HII galaxy or a
 blue compact dwarf galaxy.  All these types of dwarfs galaxies show
 evidence of starburst or post starburst activity, and are now thought to 
 harbour the majority of star formation at z $\sim$ 1. 

% As we previously discussed (section \ref{cosmologicalmodels}) at
% least two models for the creation of GRBs associate GRBs with the
% formation of massive stars: the merging of neutron star binaries
% \cite{Nara92}, and failed supernovae \cite{Woos93}.

 \begin{figure}[th]
  \centering
    \resizebox{9.0cm}{!}{\includegraphics[angle=-90]{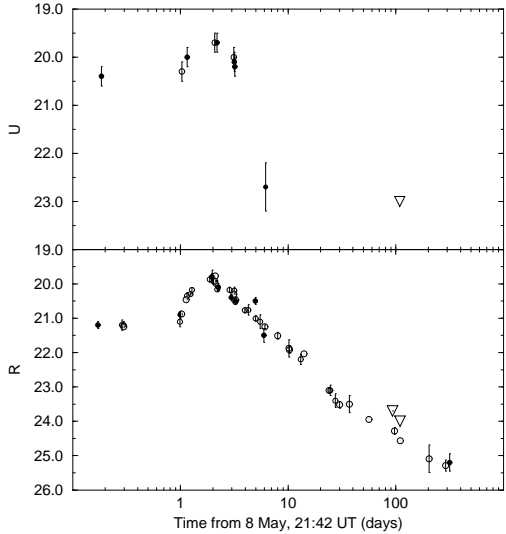}}
    \caption[The U and R-band light-curves of the optical counterpart
    of GRB 970508] {\label{970508lightcurve} The U and R-band light-curves
      of the optical transient related to GRB 970508. Filled circles are
      data based on our observations obtained at the 4.2-m WHT, 2.56-m NOT
      and 2.2-m CAHA telescopes on 9-15 May, from Castro-Tirado et al. 
      (1998b).  Empty circles are data taken from Pedersen et al. (1998), 
      Galama et al. (1998), Bloom et al. (1998a), Zharikov et al. (1998) 
      and diverse IAU circulars. Up-side down triangles are upper limits.}
 \end{figure}

\subsection{GRB 980326}

Optical observations were performed with the 2.2-m CAHA telescope (equipped 
with CAFOS). See Fig. 2.
%Due to very bad atmospherical conditions only 900 s exposures could be
%obtained. The three images were taken at a
%moderate-high airmass ($ \sec~z \sim 1.8$). 
The resulting 2700-s co-added image revealed an object  
(2.6$\sigma$ confidence level), consistent with the position of the
optical counterpart discovered by Groot et al. (1998). We derive 
R = 22.8 $\pm$ 0.7, in agreement with the measurements obtained by other 
authors.

%If no host is considered, we obtain a power law index $\alpha$ is
%$\alpha = -1.70 \pm 0.13$ with $\chi^{2}/dof = 3.45$ (long dashed line). 
If we introduce a host galaxy with R$_{c} = 25.5$ as
initially proposed by Djorgovski et al. (1998), then 
we obtain  $F_R \propto t^{-2.12 \pm 0.11}$ with $\chi^2/dof=1.32$
(point-dashed line in Fig. 2).  The value of $\alpha$ (the power-law 
exponent) is consistent with the one calculated by Groot et al. (1998) 
as $\alpha = -2.10 \pm 0.13$.

\begin{figure}[th]
   \centering
    \resizebox{10cm}{!}{\includegraphics[angle=-90]{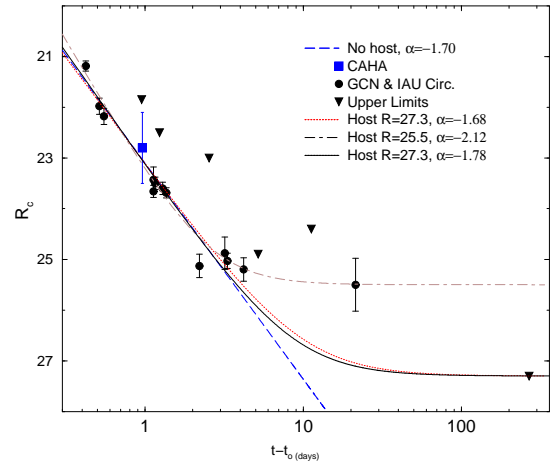}}
\caption[The R-band light curve for the GRB 980326 counterpart]
{\label{980326fig4} The R$_{c}$ band light curve for the GRB 980326
optical counterpart. The circles and triangles represent the
measurements and the upper limits taken from the literature
(GCN notices, IAU circulars) and from Groot et al. (1998). The
square is our data point obtained with the 2.2-m CAHA telescope and matches 
prety well the expected value of R$_c$ for this date. 
The point-dashed line is obtained when the data are fitted considering a 
R$_{c}$ = 25.5 host galaxy ($\alpha = -2.12$).
The dotted line is the best fit when
all the points and a host galaxy with R = 27.3 are considered ($\alpha
= -1.68$). However, this fit is not satisfactory due to the 
presence of the R$_{c} = 25.5$ data point. Once excluded, the new fit
is much better (with $\alpha$ = -1.78, solid line).
The long-dashed line is the best fit when no host galaxy is
introduced ($\alpha = -1.70$). A more extended discussion 
can be seen in Gorosabel (1999).}
\end{figure}

However, the recent upper limit of R $\geq$ 27.3 given for the 
GRB 980326 host galaxy (Bloom \& Kulkarni 1998) implies 
$F_R \propto t^{-1.70 \pm 0.13}$ with $\chi^2/dof=3.45$ (long dashed line),  
instead of $F_R \propto t^{-2.12 \pm 0.11}$ when considering a R$_c$ = 25.5 
host. In that case, the light 
curve can then be interpreted as a power-law 
decay plus an optical outburst
occurring approximately 20 days post-burst, when R$_c$ = 25.5 was 
measured for the optical transient at this date. 
This $^{\prime\prime}$secondary 
maximum$^{\prime\prime}$ could be explained in the
context of the $^{\prime\prime}$SN-like$^{\prime\prime}$ light curves 2-3 
weeks after the GRB, as recently suggested by Woosley (1999).

The value derived for the power-law decay exponent 
($\alpha$ = -1.70 $\pm$ 0.13) implies one of the fastest 
optical decay ever detected for a GRB. The existence of similar slopes in 
the GRB afterglow light curves could explain -at least in some cases- the 
non-detection of optical counterparts for many events, like GRB 970111, 
GRB 970616, GRB 970815 and GRB 970828.

\begin{acknowledgements}
We are grateful to the entire BeppoSAX team for rapidly distributing the 
GRB positions, and to the Time Allocation Committees of the German-Spanish 
Calar Alto Observatorio and Observatorio del Teide (owned by the IAC). 
We also wish to thank the referee, L. Hanlon, for 
useful suggestions, and A. Bohm, K. J. Fricke, J. Greiner and K. J\"ager 
for their help at some stage of this work. 
\end{acknowledgements}

\bibliographystyle{aa}

\end{document}